# Utility Computing and Global Grids

Chee Shin Yeo, Marcos Dias de Assunção, Jia Yu, Anthony Sulistio,
Srikumar Venugopal, Martin Placek, and Rajkumar Buyya[1]

**Gri**d Computing and **D**istributed **S**ystems (GRIDS) Laboratory
Department of Computer Science and Software Engineering
The University of Melbourne, VIC 3010, Australia
{csyeo, marcosd, jiayu, anthony, srikumar, mplac, raj}@csse.unimelb.edu.au

## 1. Introduction

*Utility computing* is envisioned to be the next generation of Information Technology (IT) evolution that depicts how computing needs of users can be fulfilled in the future IT industry [1]. Its analogy is derived from the real world where service providers maintain and supply utility services, such as electrical power, gas, and water to consumers. Consumers in turn pay service providers based on their usage. Therefore, the underlying design of utility computing is based on a service provisioning model, where users (consumers) pay providers for using computing power only when they need to.

The utility computing model offers a number of benefits to both service providers and users. From the provider's perspective, actual hardware and software components are not set up or configured to satisfy a single solution or user, as in the case of traditional computing. Instead, virtualized resources are created and assigned dynamically to various users when needed. Providers can thus reallocate resources easily and quickly to users that have the highest demands. In turn, this efficient usage of resources minimizes operational costs for providers since they are now able to serve a larger community of users without letting unused resources go unutilized. Utility computing also enables providers to achieve a better Return On Investment (ROI) such as Total Cost of Ownership (TCO) since shorter time periods are now required to derive positive returns and incremental profits can be earned with the gradual expansion of infrastructure that grows with user demands.

---

[1] Corresponding author, email – raj@csse.unimelb.edu.au

For users, the most prominent advantage of utility computing is the reduction of IT-related operational costs and complexities. Users no longer need to invest heavily or encounter difficulties in building and maintaining IT infrastructures. Computing expenditures can now be modeled as a variable cost depending on the usage patterns of users, instead of as a static cost of purchasing technologies and employing staff to manage operations. Users neither need to be concerned about possible over- or under-utilization of their own self-managed IT infrastructures during peak or non-peak usage periods, nor worry about being confined to any single vendor's proprietary technologies. With utility computing, users can obtain appropriate amounts of computing power from providers dynamically, based on their specific service needs and requirements. This is particularly useful for users who experience rapidly increasing or unpredictable computing needs. Such an outsourcing model thus provides increased flexibility and ease for users to adapt to their changing business needs and environments [2].

A study released by Meta Group in May 2004 [3] (now part of Gartner) reveals 93% of companies are seeking or adopting solutions to enable them to become an adaptive organization that is dynamically flexible in business processes and technologies (see Figure 1). Furthermore, their main goal of using these solutions is to reduce cost (see Figure 2). Companies are also worried of falling behind in competition, IT and business strategies by not becoming an adaptive organization (see Figure 3). Therefore, in today's rapidly changing business environments, there is huge market potential to employ utility computing models for companies to be more adaptive and competitive.

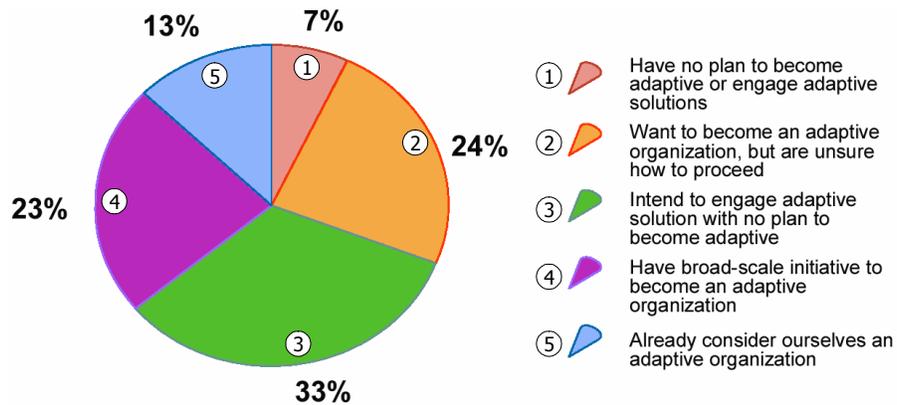

**Figure 1.** Companies' plans to become an adaptive organization (META Group [3]).

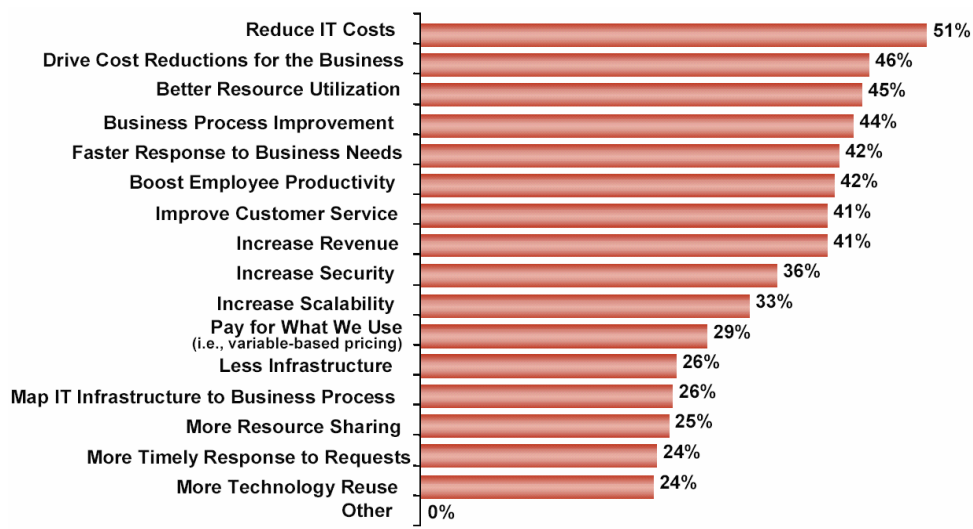

**Figure 2.** Goals of using adaptive solutions (META Group [3]).

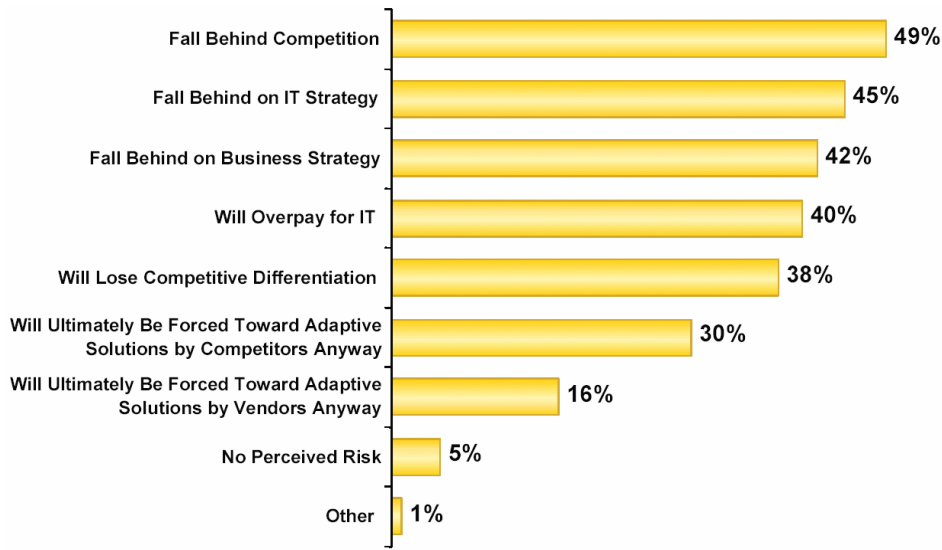

**Figure 3.** Risks of not becoming an adaptive organization (META Group [3]).

## 1.1. Potential of Grids as Utility Computing Environments

A *Grid* is an Internet-based network of geographically distributed computing resources that users can share and aggregate to solve large-scale problems [4]. An infinite number of computing devices ranging from high performance systems such as supercomputers and clusters, to specialized systems such as visualization devices, storage systems, and scientific instruments, are logically coupled together in a Grid and presented as a single unified resource [5] to the user. Figure 4 shows that a Grid user can easily use these globally distributed Grid resources by interacting with a Grid resource broker. Basically, a Grid user perceives the Grid as a single huge virtual computer that provides immense computing capabilities, identical to an Internet user who views the World Wide Web as a unified source of content.

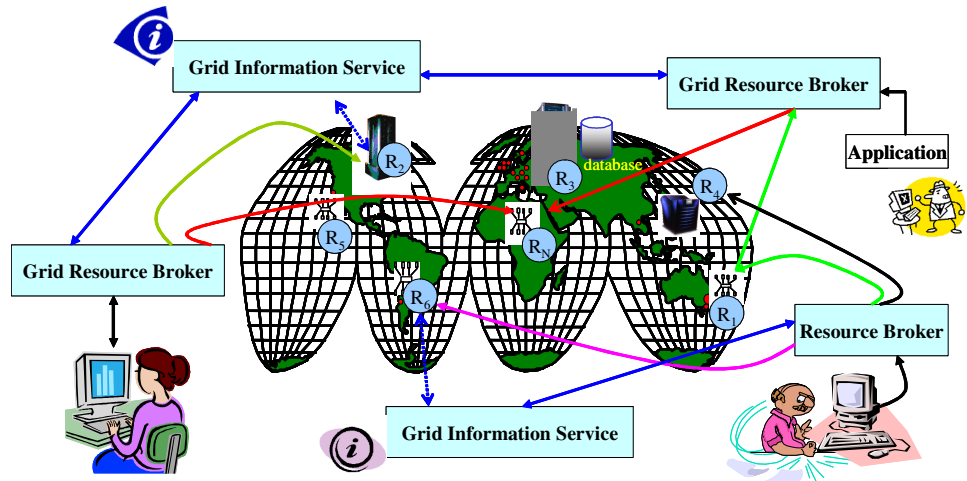

**Figure 4.** A generic view of a global Grid.

A diverse range of applications are currently or are soon to be employed on Grids, some of which include: aircraft engine diagnostics, earthquake engineering, virtual observatory, bioinformatics, drug discovery, digital image analysis, high energy physics, astrophysics, and multi-player gaming [6]. Grids can be primarily classified into three types: computational, data, and service [7]. Computational Grids aggregate computational power of globally distributed machines. Data Grids emphasize on a global-scale management of data to provide data access, integration and processing through distributed data repositories.

Service Grids (also known as Utility Grids) focus on user satisfaction by combining and delivering services based on user's needs and requirements. Accordingly, Grids are proposed as the emerging cyber infrastructure to power utility computing applications.

Grids offer a number of benefits such as:
- Transparent and instantaneous access to geographically distributed and heterogeneous resources.
- Improved productivity with reduced processing time.
- Provisioning of extra resources to solve problems that were previously unsolvable due to the lack of resources.
- A more resilient infrastructure with on-demand aggregation of resources at multiple sites to meet unforeseen resource demand.
- Seamless computing power achieved by exploiting under-utilized or unused resources that are otherwise wasted.
- Maximum utilization of computing facilities to justify IT capital investments.
- Coordinated resource sharing and problem solving through virtual organizations [8] that facilitates collaboration across physically dispersed departments and organizations.
- Service level agreement based resource allocation to meet quality of service requirements.
- Reduced administration effort with integration of resources as compared to managing multiple standalone systems.

The design aims and benefits of Grids are analogous to those of utility computing, thus highlighting the potential and suitability of Grids to be used as utility computing environments. The current trend of implementing Grids based on open standard service-based architectures to improve interoperability is a step towards supporting utility computing [9]. Even though most existing Grid applications are scientific research and collaboration projects, the number of Grid applications in business and industry-related projects is also gradually increasing. It is thus envisioned that the realization of utility computing through Grids will follow a similar course as the World Wide Web, which was first initiated as a scientific project but was later widely adopted by businesses and industries.

## 1.2. Challenges of Realizing Utility Computing Models

There are several challenges that need to be addressed in order to realize utility computing. One challenge is that both providers and users need to redraft and reorganize their current IT-related procedures and operations to include utility computing [2]. New IT policies need to be negotiated and agreed upon between providers and users, compared to the previous situation where providers and users owned and controlled their standalone policies. Providers must also understand specific service needs and requirements of users in order to design suitable policies for them. Open standards need to be established to facilitate successful adoption of utility computing so that users and producers experience fewer difficulties and complexities in integrating technologies and working together, thus reducing associated costs. Table 1 lists some of the major computing standards organizations and the activities they are engaged in.

With the changing demand of service needs from users, providers must be able to fulfill the dynamic fluctuation of peak and non-peak service demands. Service Level Agreements (SLAs) are service contracts used by providers to assure users of their level of service quality. If the expected level of service quality is not met, providers will then be liable for compensation and may incur heavy losses. Therefore, providers seek to maximize customer satisfaction by meeting service needs and minimize the risk of SLA violations [11]. Improved service-oriented policies and autonomic controls [12][13] are essential for achieving this.

Other than managing the technological aspects of delivering computing services, providers also need to consider the financial aspects of service delivery. Financial risk management for utility computing [14] is comprised of two factors: delivery risk and pricing risk. Delivery risk factors examine the risks concerned with each possible scenario in which a service can be delivered. Pricing risk factors study the risks involved with pricing the service with respect to the availability of resources. Given shorter contract durations, lower switching costs, and uncertain customer demands in utility computing environments, it is important to have dynamic and flexible pricing schemes to potentially maximize profits and minimize losses for providers [15].

**Table 1.** Some major standards organizations (J. Joseph, et al. [10]).

| Organization | Website | Standards Activities |
|---|---|---|
| Global Grid Forum (GGF) | http://www.ggf.org | Grid computing, distributed computing, and peer-to-peer networking. |
| World Wide Web Consortium (W3C) | http://www.w3c.org | World Wide Web (WWW), Extensible Markup Language (XML), web services, semantic web, mobile web, and voice browser. |
| Organization for the Advancement of Structured Information Standards (OASIS) | http://www.oasis-open.org | Electronic commerce, systems management and web services extensions, Business Process Execution Language for Web Services (BPEL4WS), and portals. |
| Web Services Interoperability Organization (WS-I) | http://www.ws-i.org | Interoperable solutions, profiles, best practices, and verification tools. |
| Distributed Management Task Force (DMTF) | http://www.dmtf.org | Systems management. |
| Internet Engineering Task Force (IETF) | http://www.ietf.org | Network standards. |
| European Computer Manufacturers Organization (ECMA) | http://www.ecma-international.org | Language standards (C++, C#). |
| International Organization for Standardization (ISO) | http://www.iso.org | Language standards (C++, C#). |
| Object Management Group (OMG) | http://www.omg.org | Model-Driven Architecture (MDA), Unified Modeling Language (UML), Common Object Resource Broker Architecture (CORBA), and real-time system modeling. |
| Java Community Process (JCP) | http://www.jcp.org | Java standards. |

This chapter focuses on the use of Grid technologies to achieve utility computing. An overview of how Grids can support utility computing is first presented through the architecture of Utility Grids. Then, utility-based resource allocation is described in detail at each level of the architecture. Finally, some industrial solutions for utility computing are discussed.

## 2. Utility Grids

Based on a market analysis conducted by Insight Research Corporation in February 2005 [16], worldwide Grid spending is projected to grow from $714.9 million in year 2005 to about $19.2 billion in 2010 as shown in Figure 5. The largest proportion of spending continues to be on enterprise Grids which is sharing resources privately within a single organization. However, there is also an exponential growth to spend on partner Grids which sharing resources among multiple organizational partners and service Grids (also known as *Utility Grids*) which offering public resource sharing globally. This analysis highlights the expected increasing adoption of Utility Grids by vendors and service providers to provide utility computing environments for businesses and industries.

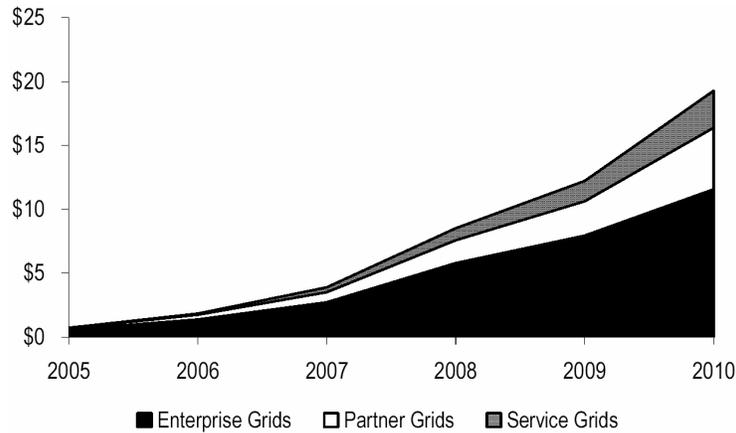

**Figure 5.** Expected worldwide Grid spending from year 2005 to 2010 in billion dollars (Insight Research Corporation [16]).

A reference service-oriented architecture for Utility Grids is shown in Figure 6. The key players in a Utility Grid are the Grid user, Grid resource broker, Grid middleware services, and Grid Service Providers (GSPs). The Grid user makes use of Utility Grids to complete their applications and thus need to express their service requirements to be fulfilled by GSPs. Given the service demand patterns and preference parameters given by the Grid user, the Grid resource broker discovers appropriate Grid middleware services and dynamically schedule applications on them at runtime, depending on their availability, capability, and costs, along with other user-defined Quality of Service (QoS) requirements. A GSP needs tools and mechanisms that support pricing specifications and schemes so they can attract users and improve resource utilization. They also require protocols that support service publication and negotiation, accounting, and payment.

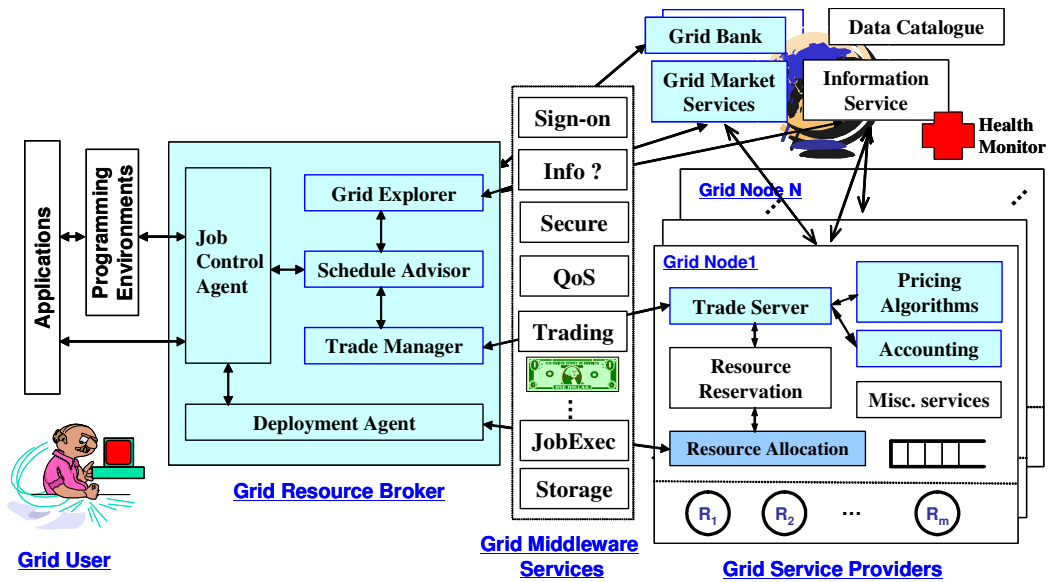

**Figure 6.** A reference service-oriented architecture for Utility Grids.

The Grid resource broker comprises the following components:
- *Job control agent*: Ensures persistency of jobs by coordinating with schedule advisor for schedule generation, handling actual creation of jobs, maintaining job status, and interacting with users, schedule advisor, and deployment agent.

- *Grid explorer*: Interacts with Grid information service to discover and identify resources and their current status.
- *Schedule advisor*: Discovers Grid resources using the Grid explorer, and select suitable Grid resources and assign jobs to them (schedule generation) to meet users' requirements.
- *Trade manager*: Accesses market directory services for service negotiation and trading with GSPs based on resource selection algorithm of schedule advisor.
- *Deployment agent*: Activates task execution on the selected resource according to schedule advisor's instruction and periodically updates the status of task execution to job control agent.

Traditional core Grid middleware focuses on providing infrastructure services for secure and uniform access to distributed resources. Supported features include security, single sign-on, remote process management, storage access, data management, and information services. An example of such middleware is the Globus toolkit [17] which is a widely adopted Grid technology in the Grid industry. Utility Grids require additional service-driven Grid middleware infrastructure that includes:
- *Grid market directory*: Allows GSPs to publish their services so as to inform and attract users.
- *Trade server*: Negotiates with Grid resource broker based on pricing algorithms set by the GSP and sells access to resources by recording resource usage details and billing the users based on the agreed pricing policy.
- *Pricing algorithms*: Specifies prices to be charged to users based on the GSP's objectives, such as maximizing profit or resource utilization at varying time and for different users.
- *Accounting and charging*: Records resource usage and bills the users based on the agreed terms negotiated between Grid resource broker and trade server.

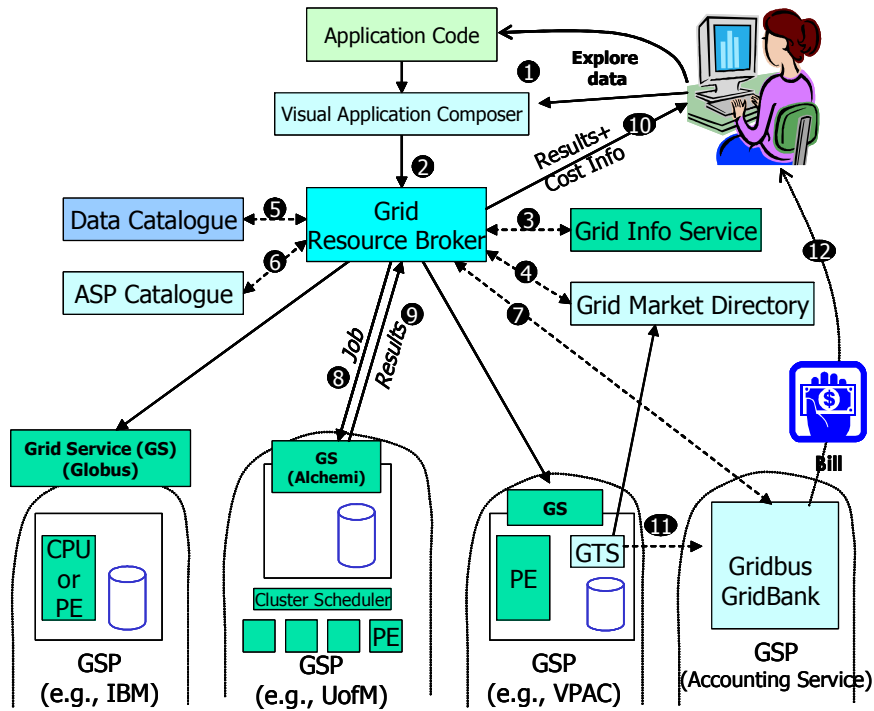

**Figure 7.** On demand assembly of services in a Utility Grid.

Figure 7 shows how services are assembled on demand in a Utility Grid. The application code is the legacy application to be run on the Utility Grid. Users first compose their application as a distributed application such as parameter sweep using visual application composer tools (Step 1). The parameter sweep model creates multiple independent jobs, each with a different parameter. This model is well suited for Grid computing environments wherein challenges such as load volatility, high network latencies, and high probability of individual node failures make it difficult to adopt a programming approach which favors tightly coupled systems. Accordingly, a parameter sweep application has been termed as a "killer application" for the Grid [18].

Visual tools allow rapid composition of applications for Grids by hiding the associated complexity from the user. The user's analysis and QoS requirements are submitted to the Grid resource broker (Step 2). The Grid resource broker first discovers suitable Grid services based on user-defined characteristics, including price, through the Grid information service and the Grid market directory (Steps 3 and 4). The broker then identifies the list of data sources or replicas through a data catalogue and selects the optimal ones (Step 5). The broker also identifies the list of GSPs that provides the required application services using the Application Service Provider (ASP) catalogue (Step 6). The broker checks that the user has the necessary credit or authorized share to utilize the requested Grid services (Step 7). The broker scheduler assigns and deploys jobs to Grid services that meet user QoS requirements (Step 8). The broker agent on the Grid resource at the GSP then executes the job and returns the results (Step 9). The broker consolidates the results before passing them back to the user (Step 10). The metering system charges the user by passing the resource usage information to the accounting service (Step 11). The accounting service reports remaining resource share allocation and credit available to the user (Step 12).

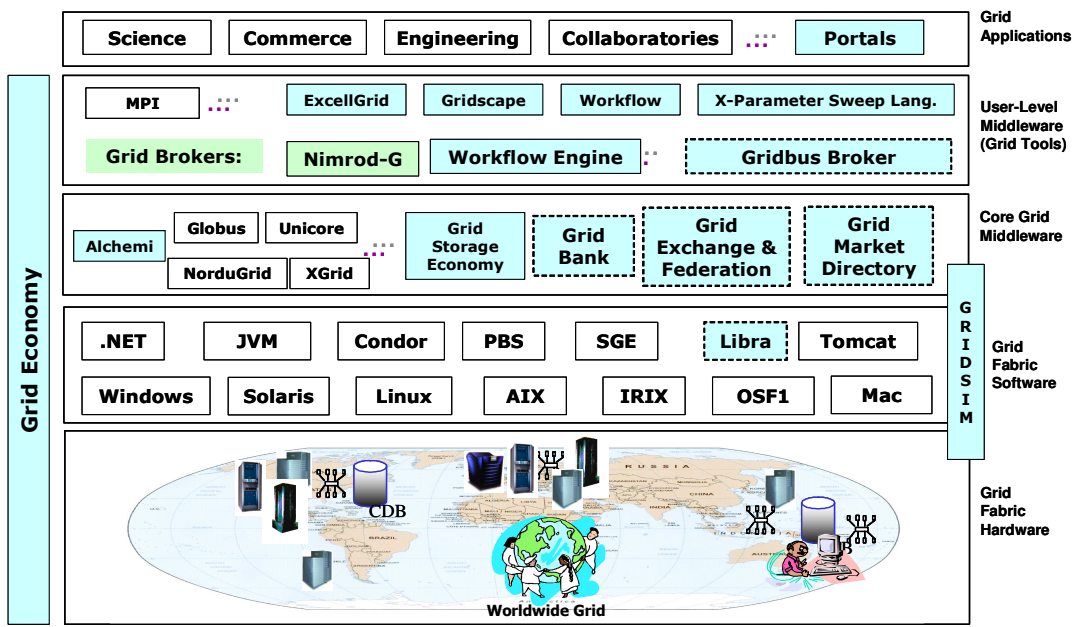

**Figure 8.** Realizing Utility Grids: Gridbus and complementary technologies.

Figure 8 shows various Grid technologies that are available to realize different layers of the Utility Grid architecture. Grid economy [19] is essential to achieve the utility model and thus has to be adopted at various layers of the architecture:
- *Grid fabric software layer*: Provides resource management and execution environment at local Grid resources.
- *Core Grid middleware layer*: Provides Grid infrastructure and essential services which consists of resource access mediator, storage, accounting, and payment.
- *User-level middleware layer*: Provides programming frameworks and policies for various types of applications and resource brokers to select appropriate and specific resources for different applications.

To enable Utility Grids, the Gridbus project has offered open-source Grid middleware [20] for various layers (as highlighted in Figure 8) that include:
- *Grid fabric software layer*: Libra, a utility-driven cluster scheduler that considers and enforces SLAs for jobs submitted into the cluster.

- *Core Grid middleware layer*: Alchemi which is a .NET-based desktop Grid framework, Grid Market Directory which is a directory publishing available Grid services, Grid Bank which provides accounting, authentication and payment facilities, and GridSim which is a event-driven simulator that models Grid environments.
- *User-level middleware layer*: Gridbus broker that selects suitable Grid services and schedules applications to run on them, Grid workflow engine that provides workflow execution and monitoring on Grids, and Visual Parametric Modeler that provides a graphical environment to parameterize applications.
- *Grid application layer*: Web portals such as Gridscape that provide interactive and dynamic web-based Grid monitoring portals and G-Monitor that manages execution of applications on Grids using brokers.

## 3. Utility-based Resource Allocation at Various Levels

Utility-based resource allocation is essential at various levels of the Utility Grid in order to realize the utility computing model. This section examines the challenges involved in clusters, distributed storage, computational Grid brokering, Data Grids, workflow scheduling, advanced reservation, and cooperative virtual organizations.

### 3.1. Clusters

*Clusters* [21] are High Performance Computing (HPC) systems that have recently been a viable choice for executing compute-intensive applications due to their offering of high performance capability at a lower cost than traditional supercomputing systems. A cluster is a collection of standalone machines that are connected via a high-speed network and installed with middleware to provide a single system image [5] to the user. An example of cluster middleware is the cluster Resource Management System (RMS) which provides a uniform interface to manage resources, queue jobs, schedule jobs and execute jobs on multiple compute nodes within the cluster.

Currently, service-oriented Grid technologies are employed to enable utility computing environments where the majority of Grid resources are clusters. Grid schedulers such as brokers and workflow engines can then discover suitable Grid resources and submit jobs to them on the behalf of the users. If the chosen Grid resource is a cluster, these Grid schedulers then interact with the cluster RMS to monitor the completion of submitted jobs. With a utility model, users can specify different levels of service needed to complete their jobs. Thus, providers and users have to negotiate and agree on SLAs that serve as contracts outlining the expected level of service performance. Providers can then be liable to compensate users for any service under-performance. So, the cluster RMS must be able to enable utility-driven cluster computing by supporting SLA based resource allocation that can meet competing user demands and enforce their service needs.

Existing cluster RMSs need to be enhanced or extended to adopt utility-driven resource allocation rather than the current system-centric resource allocation that maximizes resource throughput and utilization of the cluster. System-centric approaches assume that all job requests are equally important and thus neglect the actual levels of service required by different users. The cluster RMS should have the following components:
- *Request examiner*: Interprets the QoS parameters defined in the SLA.
- *Admission control*: Determines whether a new job request should be accepted or not. This ensures that the cluster is not overloaded with too many requests such that service performance deteriorates.
- *SLA based scheduler*: New service-oriented policies need to be incorporated to allocate resources efficiently based on service requirements.
- *Job monitor*: New measurement metrics need to be used to keep track of whether the execution progress of jobs meets the service criteria.
- *Accounting*: Maintains the actual usage of resources so that usage charges can be computed. Usage information can also be used to make better resource allocation decisions.

- *Pricing*: Formulate charges for meeting service requests. For example, requests can be charged based on submission time (peak/off-peak), pricing rates (fixed/variant), or availability of resources (supply/demand).
- *Job control*: Enforces resource assignment for executing requests to fulfill specified service needs.

The key design factors and issues for a utility-driven cluster RMS can be addressed from five perspectives [22]:
- *Market model*: Considers market concepts present in real-world human economies are can be applied for service-oriented resource allocation in clusters to deliver utility.
- *Resource model*: Addresses architectural framework and operating environments of clusters that need to be conformed.
- *Job model*: Examines attributes of jobs to ensure that various job types with distinct requirements can be fulfilled successfully.
- *Resource allocation model*: Analyzes factors that can influence the resource assignment outcome.
- *Evaluation model*: Assesses the effectiveness and efficiency of the cluster RMS in satisfying the utility model.

There is therefore growing research interest in formulating effective service-oriented resource allocation policies to satisfy the utility model. Computation-at-Risk (CaR) [23] determines the risk of completing jobs later than expected based on either the makespan (response time) or the expansion factor (slowdown) of all jobs in the cluster. Cluster-On-Demand [24] reveals the significance of balancing the reward against the risk of accepting and executing jobs, particularly in the case of unbounded penalty. QoPS [25] incorporates an admission control to guarantee the deadline of every accepted job by accepting a new job only if its deadline can be guaranteed without violating the deadlines of already accepted jobs. LibraSLA [26] accepts jobs with hard deadlines only if they can be met, and accepts jobs with soft deadlines depending on their penalties incurred for delays. Another work [27] addresses the difficulties encountered by service providers when they rent resources from resource providers to run jobs accepted from clients. These include difficulties such as which jobs to accept, when to run them, and which resources to rent to run them. It analyzes the likely impact on these difficulties in several scenarios, including changing workload, user impatience, resource availability, resource pricing, and resource uncertainty.

### 3.2. Distributed Storage

Storage plays a fundamental role in computing as it is present in many key components, from registers and Random Access Memory (RAM) to hard disk drives and optical disk drives. Combining storage with networking has created a platform for Distributed Storage Systems (DSS), and as network infrastructure continues to evolve, new exciting possibilities continue to emerge. The wide proliferation of the Internet has created a global network – a platform with endless possibilities and the source of much innovative research. DSS have the capability to function in a global environment, where storage may be shared across geographic, institutional, and administrative boundaries. The mechanism employed to make the sharing of distributed storage possible is the focus of our discussion. A key challenge is ensuring that storage services are shared fairly among users and that no single user can deny access to others, accidentally or otherwise.

One way to ensure the fair sharing of services is by applying economic principles. Examples of distributed storage systems which apply economic principles to manage various aspects operational behavior include: Mungi which manages storage quota, Mojo Nation which instills cooperative behavior, Stanford Archival Repository (SAR) which encourages the sharing of storage services and exchanges, and OceanStore which provides utility storage. SAR [28] discusses the Stanford Archival Repository, a bartering storage system for preserving information. Institutions that have common requirements and storage infrastructure can use the framework to barter with each other for storage services. For example, libraries may use this framework to replicate their archives among each other for the purpose of preservation. OceanStore [29] is a globally scalable storage utility, providing paying users with a durable, highly available storage service by utilizing non-trusted infrastructure. Mungi [30] is a Single-Address-Space Operating System (SASOS), which employs economic principles to manage storage quota. Mojo Nation [31] uses digital currency, *Mojo*, to encourage users to share resources on its network. Users that contribute services are rewarded with *Mojo*, which can then be traded for services. Storage Exchange [32] applies a Double Auction (DA) market model allowing storage services to be traded in a global environment.

Treating storage as a tradable commodity provides incentives for users to participate in the federating and sharing of their distributed storage services. However, there are still some challenges that need to be overcome before storage utility can be fully realized:
- *Federate*: A plethora of heterogeneous storage systems exist, creating a homogenous interface is a key step in federating storage.
- *Share*: Ensuring that distributed storage resource is shared fairly among users and that no single user can deny access to others, accidentally or otherwise.
- *Security*: Operating on non-trusted infrastructure requires the use of cryptographic mechanisms in order to enforce authentication and prevent malicious behavior.
- *Reliability*: Storage medium and network failures are common in a global storage infrastructure, and therefore mechanisms of remote replicas and erasure codes need to be employed to ensure that persistent reliable access to stored data is achieved.

Whilst the challenges are many, the weight of incentives and future possibilities from realizing a globally distributed storage utility ensure the continuing research and development in this area:
- *Monetary gain*: Institutions providing storage services (providers) are able to better utilize existing storage infrastructure in exchange for monetary gain. Institutions consuming these storage services (consumers) have the ability to negotiate for storage services as they require them, without needing to incur the costs associated with purchasing and maintaining storage hardware.
- *Common objectives*: There may be institutions that wish to exchange storage services between themselves due to the presence of a mutual goal such as preservation of information [28].
- *Spikes in storage requirements*: Research institutions may require temporary access to mass storage [33] such as needing access to additional storage to temporarily store data generated from experiments that are carried out infrequently. In exchange, institutions may provide access to their storage services for use by others when they are unused.
- *Donate*: Institutions may wish to donate storage services, particularly if these services are going to a noble cause.
- *Autonomic storage*: Development of a framework to support future autonomic storage systems that will allow agents to broker storage on an as needed basis.

### 3.3. Computational Grid Brokering

A *computational Grid broker* acts as an agent for the user in computational Grids. The broker hides the complexities of the Grid from the user by performing various tasks such as resource discovery, job scheduling, and job monitoring on the behalf of the user. To determine which resources to select, the broker needs to take into account various attributes from both the user perspective such as resource requirements of the application and the resource perspective such as resource architecture and configuration, resource status (available memory, disk storage, and processing power), resource availability, network bandwidth, resource workload, and historical performance.

However, the computational Grid broker needs to consider additional service-driven attributes such as QoS requirements specified by users in order to support the utility model. For example, users may specify a deadline for the completion of their application. The user may also state the maximum price to be paid for the completion. With the user's request of deadline and price for an application, the broker tries to locate the most suitable resources. Thus, during resource discovery, the broker also needs to know the costs of resources that are set by GSPs which can be obtained from a Grid market directory service. The broker must be able to negotiate with GSPs to establish an agreed price for the user, before selecting the most suitable resources with the best price based on the provided QoS. For instance, a more relaxed deadline should be able to obtain cheaper access to resources and vice-versa. So, different users often have varying prices based on their specific needs. To achieve this, we need new scheduling algorithms that take into consideration the application processing requirements, Grid resource dynamics, users' QoS requirements such as the deadline and budget, and their optimization preferences.

The Nimrod-G resource broker [34] and Gridbus Grid service broker [35] are examples of a service-oriented computational Grid brokers for parameter sweep applications. Both brokers schedule jobs based on economic principles (through the budget that the user is willing to pay) and a user-defined QoS requirement (the deadline within which the user requires the application to be completed). Nimrod-G implements four adaptive algorithms for scheduling compute-intensive parameter sweep applications:
- *Cost Optimization*: Execution time is within the specified deadline and execution cost is the cheapest.
- *Time Optimization*: Execution time is the shortest and execution cost is within the specified budget.
- *Cost-Time Optimization*: Similar to cost optimization, but if there are multiple resources with the same cost, it applies time optimization so that execution time is the shortest given the same cost.
- *Conservative Time Optimization*: Similar to cost-time optimization, but ensures that each unprocessed job in the parameter sweep application has a minimum budget-per-job.

The Gridbus broker extends cost and time optimization to schedule distributed data-intensive applications that require access to and processing of large datasets stored in distributed repositories.

### 3.4. Data Grids

Data is one of the most important entities within any IT infrastructure. Therefore, any utility computing platform must be able to provide secure, reliable and efficient management of enterprise data, and must be able to abstract the mechanisms involved. One of the key factors in the adoption of Grids as a utility computing platform is the creation of an infrastructure for storing, processing, cataloguing and sharing the ever-expanding volumes of data that are being produced by large enterprises such as scientific and commercial collaborations. This infrastructure, commonly known as *Data Grids*, provides services that allow users to discover, transfer and maintain large repositories of data. At the very minimum, a Data Grid provides a high-performance and reliable data transfer mechanism and a data replica management infrastructure. Data manipulation operations in a Data Grid are mediated through a security layer that, in addition to the facilities provided by the general Grid security services, also provides specific operations such as managing access permissions and encrypted data transfers.

In recent years, the Grid community has adopted OGSA which leverages web service technologies such as XML and SOAP to create Grid services that follow standard platform-independent mechanisms for representation, invocation and data exchange. Grid services are currently being standardized through the Grid standards organization called the Global Grid Forum (GGF). A subset of OGSA deals with providing basic interfaces called data services that describe data and the mechanisms to access it. Data services virtualize the same data by providing multiple views that are differentiated by attributes and operations. They also enable different data sources such as legacy databases, data repositories and even spreadsheets to be treated in the same manner via standard mechanisms [36].

Virtualization of data creates many possibilities for its consumption, enabled by the coupling of Grid services to enable applications such as data-intensive workflows. Already, projects such as Virtual Data Grid (VDG) [37] represent data not in terms of physical storage but as results of computational procedures. This has at least two possibilities for users: the ability to determine the provenance of data, i.e. determines how it was produced and if it were valid, and the ability to reuse the data products in future experiments where the same procedures with the same inputs are involved. Pegasus [38] is a workflow management system from the GriPhyN project which uses the VDG to reduce workflows by substituting previously generated data wherever possible. A similar procedure is followed by Storage Resource Broker (SRB) [39] which uses stored procedures in its SRB Matrix to reduce dataflow graphs. Therefore, data virtualization isolates the users from the physical Data Grid environment.

Data virtualization is an important technology for creating a data-rich utility computing environment that is able to provide its users with the following abilities:
- Seamlessly discover and use available data services
- Plan ahead for future requirements and take preemptive action
- Create dynamic applications by combining services.

In such an environment, QoS parameters associated with a data service play an important role in data service discovery. Such parameters include the size of the data, permissions associated with access and modification, available bandwidth to storage locations, and relevance. Relevance of data can be determined from the provenance data that describes the procedures used for producing the data. Planning, scheduling, and reserving resources in advance is conducted by resource brokers that take QoS parameters into account while selecting data sources and storage locations.

### 3.5. Workflow Scheduling

With the advent of Grid and application technologies, scientists and engineers are building more and more complex applications to manage and process large data sets, and execute scientific experiments on distributed resources. Such application scenarios require means for composing and executing complex workflows. *Workflows* are concerned with the automation of procedures whereby files and data are passed between participants according to a defined set of rules to achieve an overall goal [40]. A workflow management system defines, manages and executes workflows on computing resources. Imposing the workflow paradigm for application composition on Grids offers several advantages [41] such as:

- Ability to build dynamic applications which orchestrate distributed resources.
- Utilization of resources that are located in a particular domain to increase throughput or reduce execution costs.
- Execution spanning multiple administrative domains to obtain specific processing capabilities.
- Integration of multiple teams involved in managing different parts of the experiment workflow, thus promoting inter-organizational collaborations.

QoS support in workflow management is required by many workflow applications. For example, a workflow application for maxillo-facial surgery planning [42] needs results to be delivered before a certain time. However, QoS requirements cannot be guaranteed in a conventional Grid where resources provide only best effort services. Therefore, there is a need to have Utility Grids that allow users to negotiate with service providers on a certain service agreement with the requested QoS.

In general, scheduling workflows on Utility Grids is guided by users' QoS expectations. Workflow management systems are required to allow the user to specify their requirements, along with the descriptions of tasks and their dependencies using the workflow specification. In general, QoS constraints express the preferences of users and are essential for efficient resource allocation. QoS constraints can be classified into five dimensions [43]: *time*, *cost*, *fidelity*, *reliability* and *security*. Time is a basic measure of performance. For workflow systems, it refers to the total time required for completing the execution of a workflow. Cost represents the cost associated with the execution of workflows, including the cost of managing workflow systems and usage charge of Grid resources for processing workflow tasks. Fidelity refers to the measurement related to the quality of the output of workflow execution. Reliability is related to the number of failures for execution of workflows. Security refers to confidentiality of the execution of workflow tasks and trustworthiness of resources.

Several new issues arise from scheduling QoS constrained workflows on Utility Grids:
- In general, users would like to specify a QoS constraint for the entire workflow. It is required that the scheduler determines a QoS constraint for each task in the workflow, such that the global QoS is satisfied.
- The scheduler is required to be adaptive to evaluate a proposed SLA and negotiate with a service provider for one task with respect to its current accepted set of SLAs and expected return of unscheduled tasks.
- The description language and monitoring mechanism for QoS based workflows will be more complex as compared to traditional workflow scheduling.

To date, several efforts have been made towards QoS aware workflow management. Grid Quality of Service Management (G-QoSm) [44] provides Grid services which workflow schedulers can negotiate and reserve services based on certain quality levels. The Vienna Grid Environment (VGE) [45] develops a dynamic negotiation model that facilitates workflow schedulers to negotiate various QoS constraints with multiple service providers. It also extends the Business Process Execution Language (BPEL) [46] to support QoS constraint expression. A QoS based heuristic for scheduling workflow applications can be found in [47]. The heuristic attempts to assign the task into least expensive computing resources based on assigned time constraint for the local task. A cost based scheduling algorithm [48] minimizes the cost, while meeting the deadline of the workflow by distributing the deadline for the entire workflow into sub-deadlines for each task. More recently, a budget constrained workflow scheduling has been developed in [49]. It uses genetic algorithms to optimize workflow execution time while meeting the user's budget.

However, supporting QoS in scheduling of workflow applications is still at a very preliminary stage. There is a need for many advanced capabilities in workflow systems and resource allocation such as support of cyclic and conditional checking of workflow structure, adaptive scheduling based on dynamic negotiation models, and advanced SLA monitoring and renegotiation.

### 3.6. Advanced Reservation

In existing Grid systems, incoming jobs to resources are either scheduled via Space-shared or Time-shared mode. The Space-shared mode runs jobs based on their submission times; similar to First Come First Serve (FCFS), whereas the Time-shared mode allows multiple executions of jobs; hence it behaves like a Round Robin approach. With a Utility Grid, jobs can be prioritized based on users' QoS by a resource scheduler. However, a Utility Grid is not necessarily able to handle high priority jobs or guarantee reliable service. Therefore, *advance reservation* needs to be introduced in a Utility Grid system to secure resources prior to their execution.

Advanced Reservation (AR) is a process of requesting resources for use at a specific time in the future [50]. Common resources that can be reserved or requested are processors, memory, disk space and network bandwidth or a combination of any of those. The main advantage of AR is that it guarantees the availability of resources to users and applications at specific times in the future. Hence, from the user's perspective:
- Jobs can be executed straight away at a specified time rather than being held up in a queue by other jobs. This is a highly-desirable approach for executing workflow applications that have one or more dependencies.
- Avoiding any dropouts in a network transfer which is not an option in multimedia streaming applications such as video-conferencing.

Combining utility computing with AR allows resource providers to specify criteria and requirements of usage. In addition, resource providers can match and satisfy user's QoS. Utility computing applies to different stages of AR [51] as follows:
- *Requesting a new reservation slot*: A user asks a resource about the availability of a reservation slot by giving details such as start time, duration time, and number of processors. A resource can then either accept or reject the request (if the slot is already booked). Both the user and resource can continue negotiating until they reach an agreement, i.e. to accept or not to accept.
- *Modifying an existing reservation slot*: A user or a resource can request to modify an existing reservation slot. Modification is used as a way to shorten or extend the duration of a reservation. If a user's jobs are running longer than expected, extending the duration time is needed to prevent them from being preempted by a resource. A user can request to shorten a reservation if jobs have finished, in order to save some costs.
- *Canceling an existing reservation slot*: A user or a resource can request to cancel an existing reservation slot. However, this can be done only when both parties have agreed beforehand. Canceling a reservation before it starts can easily be agreed to by a resource. However, some resource providers may not allow the cancellation of a reservation once it has started.

### 3.7. Cooperative Virtual Organizations

The concept of *virtual organization* (VO) is crucial to the Grid. In current Grid collaborations, physical organizations engage in projects and alliances such as joint ventures that require shared access to compute and data resources provided by their members. The model adopted for such endeavors is that of a VO, in which resource providers and users are organized in a structure that may comprise several physical organizations. VOs may vary in several ways, such as scope, dynamism, and purpose, even though they impose similar challenges regarding their formation, operation and dissolution [52].

Over the years, applications and compute and data resources have been virtualized to enable the utility model for business processes. With regard to the creation of a VO, it may be assumed that a VO is formed because of the need from a business process or some project. Despite the virtualization provided by Grid technologies, organizations may have difficulties in expressing their needs and requirements to their potential partners. Additional challenges are the selection of partners and the establishment of trust at a level that allows the automated creation of VOs [53]. However, to enable the utility model in VOs, issues regarding the responsive or even the automated creation of VOs need to be tackled.

The operational phase of a VO is also a complex task. Resource sharing in VOs is conditional and rules-driven. Also, the relationship in some VOs is peer-to-peer. To complicate matters further, the collection of participating entities is dynamic [10]. This scenario complicates tasks such as the negotiation of SLAs among the participants of the VO or between the participants and the VO itself. Furthermore, the reconciliation, management, and enforcement of resource usage control policies in the VO poses several challenges as presented in [54]. For example, a simple model for providing resources as utilities in VOs requires the presence of a trusted VO manager. Resource providers are committed to deliver services to the VO according to contracts established with the VO manager. The manager is therefore responsible for assigning quotas of these resources to VO groups and users based on some VO policy. Users are allowed to use services according to these quotas and the VO policy. However, in this context, the delivery of compute and data resources in a utility-model to VO users and groups makes tasks such as enforcement of policies in a VO level and accounting difficult.

A simple model in a VO that follows a peer-to-peer sharing approach is of best effort, in which "you give what you can and get what others can offer". A more elaborate model in which the presence of a VO manager does not exist allows the delivery of services following a "you get what give" approach [55]. Other approaches require multilateral agreements among the members of the VO and give rise to challenges in the enforcement of resource usage policies, as described before.

Current works have not focused on aspects related to the dissolution of VOs, since this problem involves more legal and social issues rather than technical issues. Hence, the delivery of compute resources as a utility in VOs requires the investigation and solving of these problems related to various aspects of a VO. Automation and responsiveness are also required in every stage of the lifecycle of a VO.

## 4. Industrial Solutions for Utility Computing

Various commercial vendors have launched industrial solutions to support utility computing. Competing marketing terms are used by different vendors even though they share the same vision of providing utility computing. This section discusses three major industrial solutions, as listed in Table 2: HP's Adaptive Enterprise, IBM's E-Business on Demand, and Sun Microsystems's Sun Grid. All three solutions use Grids as the core enabling technology.

### 4.1. HP Adaptive Enterprise

The vision of HP's *Adaptive Enterprise* [56] is to synchronize the business and IT processes in an enterprise in order to allow it to benefit from changes in market demands. IT processes are coordinated through the Adaptive Enterprise architecture which comprises two dimensions: IT management and IT service capabilities.

The IT management dimension involves:
- *IT business management*: Long-term IT strategies such as asset management, customer and supplier relationship management, and project portfolio management need to be developed.
- *Service delivery management*: Various operational aspects of service delivery such as availability, cost, capacity, performance, security, and quality need to be considered.
- *Service delivery*: IT services need to be provided by highly automated systems to users.

The IT service capabilities dimension that is addressed under both service delivery management and service delivery in the IT management dimension consists of:
- *Business services*: Represent top-level services related to business processes such as the composition of workflows.

- *Information services*: Consolidate and manipulate information for business services that are independent from application services.
- *Application services*: Automate and handle the processing of applications.
- *Infrastructure services*: Create a common infrastructure platform to host the application, information, and business services.

**Table 2.** Some major industrial solutions for utility computing.

| Vendor | Solution and Website | Brief Description | Core Enabling Technology and Website |
|---|---|---|---|
| HP | Adaptive Enterprise<br>http://www.hp.com/go/adaptive | Simplifies, standardizes, modularizes, and integrates business processes and applications with IT infrastructures to adapt effectively in a changing business environment. | Grids<br>http://www.hp.com/go/grid |
| IBM | E-Business On Demand<br>http://www.ibm.com/ondemand | Performs on demand business processes that include research and development, engineering and product design, business analytics, and enterprise optimization. | Grids<br>http://www.ibm.com/grid |
| Sun Microsystems | Sun Grid<br>http://www.sun.com/service/sungrid | Offers computing power pay-as-you-go service utility by charging users $1 for every hour of processing. | Grids<br>http://www.sun.com/software/grid |

The Adaptive Enterprise architecture also defines four design principles that are to be realized for an enterprise to become more adaptive:
- *Simplification*: Complex IT environments can be streamlined through application integration, process automation, and resource virtualization to facilitate easier management and faster response.
- *Standardization*: Standardized architectures and processes enable easier incorporation of new technologies, improves collaboration and saves cost.
- *Modularity*: Smaller reusable components can be deployed faster and more easily, increasing resource sharing and reducing cost.
- *Integration*: Dynamic linking of business processes, applications, and infrastructure components enhance agility and cost efficiency.

Grid technologies enable the successful implementation of the Adaptive Enterprise architecture to link IT infrastructure dynamically to business processes by fulfilling the following design rules:
- *Service-oriented architecture (SOA)*: Grid services are defined based on the Open Grid Service Architecture (OGSA) [57], an open standard that leverages web services to allow large-scale collaboration across the Internet.
- *Virtualization*: Grids harness a large pool of resources that can be shared across applications and processes to meet business demands.
- *Model-based automation*: Grid technologies integrate standalone resources and automate services, thus simplifying the process of deployment, management, and maintenance.

HP customizes the Globus Toolkit [17] as the Grid infrastructure for its platforms. The Grid solutions offered by HP and its partners are listed in Table 3.

**Table 3.** HP and its partners' Grid solutions.

| HP Solutions and Website | HP Partner Solutions and Website |
|---|---|
| - Management Solutions: *OpenView* <br> http://www.hp.com/go/openview <br> - Server Solutions: *BladeSystem* <br> http://www.hp.com/go/bladesystem <br> - Storage Solutions: *StorageWorks Grid* <br> http://www.hp.com/go/storageworksgrid | Infrastructure <br> - Application Infrastructure: *DataSynapse* <br> http://www.datasynapse.com <br> - Grid Infrastructure: *United Devices* <br> http://www.ud.com <br> - Resource Management: *Axceleon* <br> http://www.axceleon.com <br> - Workflow Management: *TurboWorx* <br> http://www.turboworx.com <br> - Workload Management: *PBS Pro* <br> http://www.altair.com <br> - Workload Management: *Platform* <br> http://www.platform.com |

## 4.2. IBM E-Business On Demand

IBM's *E-business On Demand* [58] aims to improve the competitiveness and responsiveness of businesses through continuous innovation in products and services. To achieve this aim, E-business On Demand optimizes the following business processes:
- *Research and development*: New innovative products and services need to be researched and developed quickly to make an impact in the highly competitive market.
- *Engineering and product design*: Products and services need to be well-engineered and designed to meet customers' requirements.
- *Business analytics*: Swift and accurate business decisions need to be made based on market performance data in order to remain a market leader.
- *Enterprise optimization*: Standalone resources at various global branches need to be integrated so that workload can be distributed evenly and resources utilized fully to satisfy demand.

E-business On Demand targets numerous industries that include:
- *Automotive and aerospace*: Collaborative design and data-intensive testing.
- *Financial services*: Complex scenario simulation and decision-making.
- *Government*: Coordinated operation across civil and military divisions and agencies.
- *Higher education*: Advanced compute and data-intensive research.
- *Life sciences*: Biological and chemical information analysis and decoding.

IBM uses Grid technologies as the key component to provide the flexibility and efficiency required for various E-Business On Demand environments:
- *Research and development*: Highly compute- and data-intensive research problems can be solved with lower cost and shorter time by harnessing extra computational and data resources in a Grid.
- *Engineering and product design*: Industry partners are able to collaborate by sharing resources and coordinating engineering and design processes through VOs and open standards-based Grid architecture.
- *Business analytics*: Heavy data analysis and processing can be sped up with extra computational and data resources so that results are derived in time for decision-making.
- *Enterprise optimization*: Virtualization and replication using Grid technologies ensures that under-utilized resources are not wasted, but are instead utilized for backup and recovery purposes.

The Grid infrastructure that IBM deploys is called the IBM Grid Toolbox which is an enhanced version of the Globus Toolkit [17]. Table 4 lists Grid solutions that are available by IBM and its partners. IBM provides a general integrated Grid solution offering called Grid and Grow for interested customers to easily deploy and sample Grid technologies. In addition, it has created customized Grid offerings for specific industries and applications to drive E-business On Demand.

**Table 4.** IBM and its partners' Grid solutions.

| IBM Solutions and Website | IBM Partner Solutions and Website |
|---|---|
| - Application Server: *Websphere* http://www.ibm.com/websphere<br>- Resource Provisioning: *Tivoli* http://www.ibm.com/tivoli<br>- System Server: *eServer* http://www.ibm.com/eserver | Infrastructure<br>- Application Infrastructure: *DataSynapse* http://www.datasynapse.com<br>- Grid Infrastructure: *United Devices* http://www.ud.com<br>- Grid Infrastructure: *Univa* http://www.univa.com<br>- Workload Management: *PBS Pro* http://www.altair.com<br>- Workload Management: *Platform* http://www.platform.com<br><br>Application<br>- Document Production: *Sefas* http://www.sefas.com<br>- Grid Deployment: *SAS* http://www.sas.com/grid<br>- Risk Management: *Searchspace* http://www.searchspace.com |

### 4.3. Sun Microsystems Sun Grid

Sun Microsystems's *Sun Grid* aims to provide affordable commodity-based computing power pay-as-you-go service. Sun Grid currently supports three types of compute utility service (see Table 5 for comparison):
- *Compute utility*: A standard offering that provides instant deployment for anyone with Internet access at $1 per hour of processing.
- *Commercial utility*: A single tenant standard offering in a multi tenant hosting center that targets medium and large enterprises.
- *Variable cost infrastructure*: A customized modular offering and single tenant utility model for large enterprises and system integrators.

The Sun Grid compute utility [59] provides a golden opportunity for non-IT users to make use of utility computing by providing a simple and easy to use web interface that hides the complex Grid technologies involved. Given this assumption of a simple utility computing environment, the Sun Grid compute utility has several limitations:
- Submitted applications must be able to execute on Solaris 10 operating system.
- Submitted applications must be self-contained and scripted to work with Sun N1 Grid Engine software without requiring interactive access.
- Submitted applications has to be implemented using standard object libraries included with Solaris 10 Operating system or user libraries packaged with the executable.
- The application can obtain finer control over compute resources through the interfaces provided by the Sun N1 Grid Engine software.
- The total maximum size of applications and data must be less than 10 GBytes.

- Applications and data can only be uploaded through the web interface and may be packaged into compressed ZIP files of less than 100 MBytes each.

**Table 5.** Comparison of Sun Grid compute utility services.

| Comparison | Sun Grid Compute Utility Services | | |
| --- | --- | --- | --- |
| | Compute Utility | Commercial Utility | Variable Cost Infrastructure |
| Access | Portal | Dedicated | Customer or system integrator hosted |
| Availability | Instantaneous | Short notice | Longer term contract |
| Scheduling | No reservation | Time-based reservation | Dedicated or time-based reservation |
| Pricing | All-inclusive $1/CPU-hour | Negotiated $/CPU-hour | Negotiated |
| Business Terms | Standard | Service level | Capacity provisioning |
| Technology | Solaris 10 x64 | Solaris 10 x64 or Redhat Linux | Menu of options |
| Storage | 10 GB standard | Customer defined | Menu of options |

**Table 6.** Sun Microsystems and its partners' Grid solutions.

| Sun Microsystems Solutions and Website | Sun Microsystems Partner Solutions and Website |
| --- | --- |
| - Resource Management: *N1* <br> http://www.sun.com/software/n1gridsystem | Infrastructure <br> - Application Server: *GigaSpaces* <br> http://www.gigaspaces.com <br> - Autonomic Processing: *Paremus* <br> http://www.paremus.com <br> - Workload Management: *Platform* <br> http://www.platform.com <br><br> Service Management <br> - Collaborative Solutions: *SAP* <br> http://www.sap.com <br> - Database Solutions: *Oracle* <br> http://www.oracle.com <br> - Service-Oriented Solutions: *BEA* <br> http://www.bea.com <br><br> Software As a Service <br> - Pricing and Risk Solutions: *CDO2* <br> http://www.cdo2.com |

Grid technologies are employed for the Sun Grid compute utility with the following compute node configuration:
- 8 GBytes of memory.
- Solaris 10 operating system.

- Sun N1 Grid Engine 6 software for resource management of compute nodes.
- Grid network infrastructure built on Gigabit Ethernet.
- Web-based portal for users to submit jobs and upload data.
- 10 GBytes of storage space for each user.

Industries that are targeted by the Sun Grid compute utility consists of:
- *Energy*: Reservoir simulations and seismic processing.
- *Entertainment/Media*: Digital content creation, animation, rendering, and digital asset management.
- *Financial services*: Risk analysis and Monte Carlo simulations.
- *Government education*: Weather analysis and image processing.
- *Health sciences*: Medical imaging, bioinformatics, and drug development simulations.
- *Manufacturing*: Electronic design automation, mechanical computer-aided design, computational fluid dynamics, crash-test simulations, and aerodynamic modeling.

Sun Microsystems and the Globus project have been collaborating in the joint development of interfaces between Sun Grid solutions and the Globus Toolkit [17]. Table 6 shows Grid solutions that are developed by Sun Microsystems and its partners.

## 5. Summary

In this chapter, the utility computing model and its vision of being the next generation of IT evolution is introduced. The utility computing model is significantly different from traditional IT models, and thus requires organizations to amend their existing IT procedures and operations towards this outsourcing model so as to save costs and improve quality. There is also increasing emphasis on adopting Grid computing technologies to enable utility computing environments.

This chapter has focused on the potential of Grids as utility computing environments. A reference service-oriented architecture of Utility Grids has been discussed, along with how services are assembled on demand in the Utility Grid. The challenges involved in utility-based resource allocation at various levels of the Utility Grid are then examined in detail. With commercial vendors rapidly launching utility computing solutions, industrial solutions by three pioneer vendors (HP, IBM, and Sun Microsystems) and their realization through Grid technologies are also presented.

For recent advances in Grid computing technologies and applications, readers are recommended to browse the proceedings of CCGrid [60], Grid [61], and e-Science [62] conference series organized by the IEEE Technical Committee on Scalable Computing (TCSC) [63].

## 6. Acknowledgements
The authors thank Hussein Gibbins for his comments. This work is partially supported through the Australian Research Council (ARC) Discovery Project grant.## 7. Glossary

*Adaptive Enterprise* – An organization that is able to adjust and benefit according to the changes in its operating environment.

*Grid Computing* – A model allowing organizations to access a large quantity of remotely distributed computing resources on demand.

*Market-based Resource Allocation* – Assignment of resources based on market supply and demand from providers and users.

*Middleware* – Software designed to interface and link separate software and/or hardware.

*On Demand Computing* – Computing services that can be accessed when required by the user.

*Service Level Agreement (SLA)* – A contract agreed upon between a service provider and a user which formally specifies service quality that the provider is required to provide.

*Service-Oriented Architecture (SOA)* – An architectural framework for the definition of services that are able to fulfill the requirements of users.

*Utility Computing* – A model whereby service providers offer computing resources to users only when the users need them and charges the users based on usage.

*Virtual Organization (VO)* – A temporary arrangement formed across physically dispersed departments and organizations with a common objective to facilitate collaboration and coordination.